# INJECTION CONTROL SYSTEM OF HLS STORAGE RING

Gongfa Liu, Jingyi Li, Weimin Li, Chuan Li, Kaihong Li, Lei Shang
National Synchrotron Radiation Lab., P. O. Box 6022, Hefei, Anhui 230029, P. R. China

*Abstract*

The injection control system of Hefei Light Source (HLS) storage ring is a subsystem of the upgraded HLS control system, which is based upon EPICS. Three programmable logic controllers (PLCs) are used as device controllers, which control one septum modulator and four kicker modulators of HLS storage ring. An Industrial PC is used as Input/Output Controller (IOC) and it connects the PLCs with serial communication (RS232 mode) over fibre. A PC with Linux is used as operator interface (OPI), operator application are running on it. The control system was completed in July 2000. The commissioning shows that the control system is reliable and easy operational.

## 1 INTRODUCTION

HLS is a dedicated synchrotron light source in which the electrons circulate at the energy of 800Mev in the storage ring. As a part of NSRL Phase II project, the injection system is updated, four kicker mode substitutes for three kicker mode. Three programmable logic controllers (PLCs) are used as device controllers, which control one septum modulator and four kicker modulators of HLS storage ring. An Industrial PC is used as Input/Output Controller (IOC) and it connects the PLCs with serial communication (RS232 mode) over fiber. Strong electromagnetic interference, which is caused by the big pulse current in the modulator, is restrained. A PC with Linux is used as operator interface (OPI), operator application are running on it. The control system was completed in July 2000. The commissioning shows that the control system is reliable and easy operational.

## 2 HARDWARE CONFIGUREATIONS

The injection control system of HLS storage ring is built upon EPICS. It is a subsystem of the HLS control system. Fig.1 shows the structure of the storage ring injection control system.

We use a linux PC with a Pentium III 500 CPU as the OPI. The man-machine interface shows on it. An IPC with a Pentium II 350 CPU is used as IOC. In order to provide high-speed communication between OPIs and IOCs, we use 100M Ethernet at the network level. An Intel Pro/100+ Ethernet card was added to the IOC crate to provide an interface between the IOC and the network. Three PLCs (Omron C200H) are used as device controllers, one for septum modulator, the other two for four kicker modulators. Each PLC has an Optical Host Link Unit. They are connected in series using optical fiber cable. Link Adapter (3G2A9-AL004-(P)E) is used as converter between RS232 cable and optical fiber cable.[1]

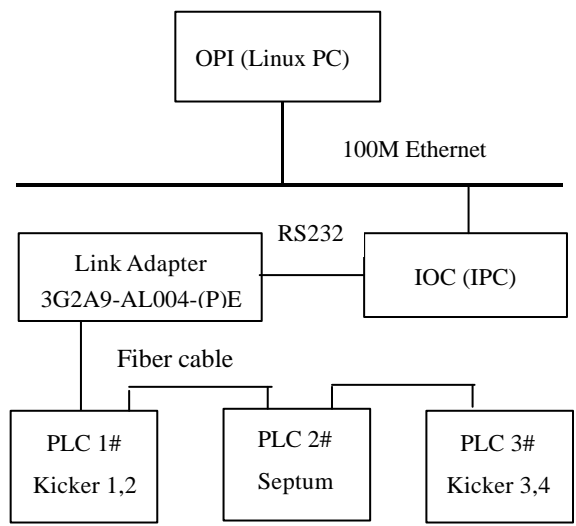

Fig.1 Structure of storage ring injection control system.

## 3 SOFTWARE OVERVIEW

### 3.1 Software of PLC

We use ladder software package LSS to program. The following function are achieved: remote/local control choice, low voltage power on/off, high voltage power on/off, reset, time delay, analog output set, input regulator motor control, safety interlock and communication with IOC [2].

### 3.2 Software of IOC

The heart of each IOC is a memory resident database together with various memory resident structures describing the contents of the database. The following record types are mainly used: ai(Analog Input), ao(Analog Output), bo(Binary Output), MbbiD (Direct Multi-bit Binary Iutput). DCT313 (Database Configuration Tool) is used to create a run time database for the IOC, then the Device Support and Device Drivers (if necessary) are written for each kind of record [3].

Omron PLC (C200H) provides serial communications protocol. The protocol has its Command/Response format and command set. The user's application must conform to the protocol. We write a device support and device driver under the protocol.

## 3.3 Software of OPI

We use a PC with Linux as OPI, operator application are running on it. We use MEDM to develop the man-machine interface. Fig.2 is the screen of HLS storage ring injection control system.

The operation sequence is as follow: turn on low voltage power; waiting for 15 minutes (for hyratron-switch filament warm-up); when the ready lamp is on, turn on high voltage power and trigger (for timing synchronization); then set current output. The high voltage value and discharge counts are displayed. When error lamp is on, we can press the button near the error lamp, and the error details are shown in the relational screen.

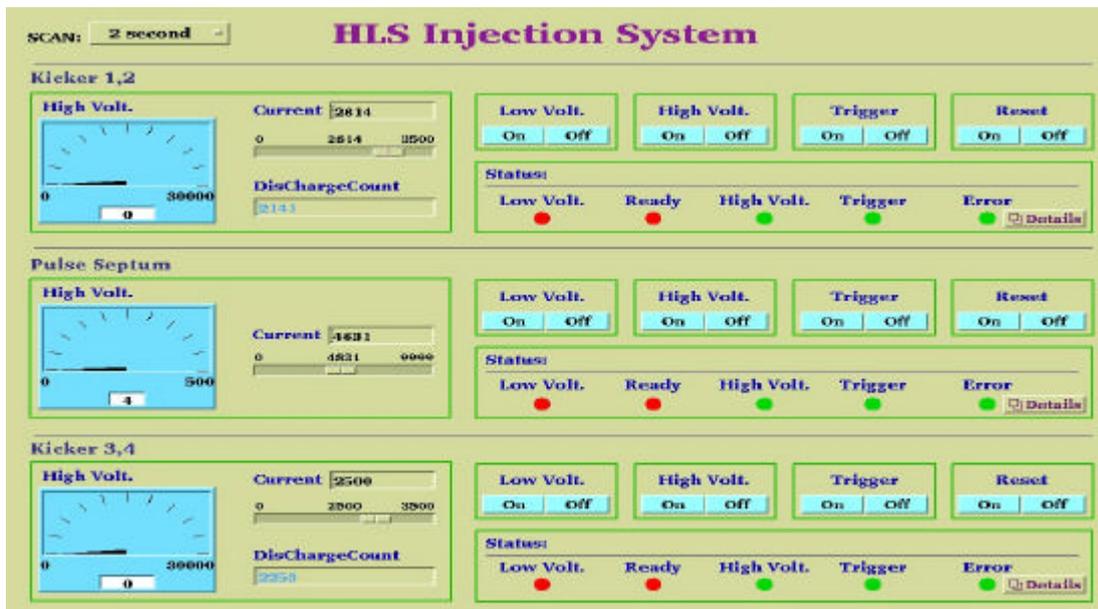

Fig. 2 the screen of HLS storage ring injection system

## 4 CONCLUSION

The injection control system of HLS storage ring is a subsystem of the upgraded HLS control system. We have finished the development of the control system of main magnet power supplies. Due to the good extensibility of EPICS, the development for HLS storage ring injection control system is easy. The control system was completed in July 2000. PLC applies industrial standard and a number of measures are taken to avoid communication error, so the system is reliable. The commissioning shows that the system is also easy operational.